# Initial Development of MBE-Grown InAs Diodes for Thermoradiative Energy Harvesting

I. Artacho[(1)], I. Ramiro[(2)] and A. Martí[(1)]

(1) Instituto de Energía Solar, ETSI Telecomunicación, Universidad Politécnica de Madrid,

Ciudad Universitaria sn, 28040 Madrid, Spain, EU

(2) Electronic Engineering Department, Universitat Politècnica de Catalunya,

Jordi Girona 31, Barcelona, 08034, Spain, EU

Abstract: We describe the development of 1×1 $mm^2$ InAs thermoradiative diodes grown by molecular beam epitaxy with emphasis on their reverse saturation current and break-down voltage. P-i-n diode structures grown at 450 C, with $As_2$ flux around 3 times stoichiometry and an In effusion cell tip temperature 150 C higher than the base temperature, exhibit the best results with breakdown voltages above 0.3 V and reverse saturation current densities 200 times the radiative limit.

## 1. Introduction

Indium arsenide (InAs) is characterized by its low direct bandgap (0.35 eV at room temperature, [1]) and high electron mobility (above 30,000 $cm^2 \cdot V^{-1} \cdot s^{-1}$, [2]). With these parameters, research into the development of InAs diodes was triggered in the past by their possible application in high-frequency, low-power dissipation electronics [3] as well as in magnetic Hall effect sensors [4], mid-infrared (MIR) photodetectors [5] and thermophotovoltaic (TPV) solar cells [6], [7]. Our interest for developing InAs diodes relies in its possible application as thermoradiative (TR) diodes to obtain electric energy from low temperature heat sources [8]. This application is briefly revisited in section 2. In section 3 we describe the InAs diode structures grown by MBE to explore this technology and discuss the results. In section IV we summarize our conclusions and point at future lines of research development.

## 2. Thermoradiative diodes

TR diodes are the counterpart of solar cells understood as thermal engines. A solar cell converts into electricity part of the radiation energy emitted towards it by a hotter body. A TR diode, instead, converts into electricity part of the energy that it would otherwise radiate to a colder environment. It is accepted that the limiting efficiency of photovoltaic single gap solar cells (40.7 % at maximum sunlight concentration) is governed by the Schockley and Queisser (S&Q) detailed balance limit [9]. In addition to the S&Q original work, additional subtle details of this theory, such as the role of photon recycling, stimulated photon emission and the apparent paradox that total photon absorption does not always lead to maximum efficiency, can be found in previous reviews by us [10], [11]. As a short reminder we will mention that, when operated at “one sun”, a solar cell is assumed to receive direct photons from the sun only from a relatively small solar semi-angle (the one corresponding to the solar dish, around 0.267º) while it emits radiative photons in all directions; in contrast, at maximum concentrated sunlight, the solar dish view is amplified and the solar cell receives photons form all directions. This mode of illumination is of interest because it is also the one at which TR diodes usually operate since, as we will see, they receive ambient photons from all directions.

Beyond these details, the practical fact relevant to this work is that, to approach this limit, radiative electron–hole pair recombination (the mechanism by which an electron recombines from the conduction band to the valence band emitting a photon) must be the dominant recombination mechanism in the diodes. When radiative recombination is dominant, according to detailed balance, the idealized current density–voltage characteristic of a solar cell at maximum concentration is given by:

$$J = \frac{2\pi}{h^3 c^2} \int_{E_G}^{\infty} \left( \frac{\epsilon^2}{\exp \frac{\epsilon}{k_B T_S} - 1} - \frac{(n^2+1)\epsilon^2}{\exp \frac{\epsilon - eV}{k_B T_C} - 1} \right) d\epsilon \quad (1)$$

where: $V$ is the operation voltage of the cell, $E_G$ is the gap of the semiconductor the solar cell is made of, $h$ is the Planck's constant, $k_B$ is the Boltzman's constant, $c$ is the speed of light in vacuum, $e$ is the proton charge, $T_C$ is the temperature of the cell and $n$ is the refraction index of the substrate the pn junction is manufactured on (when it is assumed to be sufficiently thick as to absorb all the light the pn junction emits towards it). If a back reflector is placed instead of a substrate at the rear of the diode, then we must set $n = 0$. In addition: illumination from the sun has been simplified to that of a blackbody at temperature $T_S$; the cell is assumed to absorb all photons above the energy gap and, as it is customary for obtaining maximum efficiency, full solar concentration is assumed.

To gain insight into the TR diode mode of operation, Fig. 1 illustrates (black curve) the ideal current density – voltage characteristic of an ideal InAs diode operating in the radiative limit, with back reflector ($n = 0$) at $T_C = 300$ K in dark conditions (that is, in equilibrium with the ambient at $T_S = 300$ K). TR operation, with the diode at $T_C$ =550 K and the ambient at $T_S = 300$ K, is represented by the blue curve. The maximum power delivered in this mode of operation for this example is 13 W·m$^{-2}$. For reference, we also plot (red curve) the same diode operating in photovoltaic model (TPV operation) with at $T_C$ =300 K and the ambient at $T_S = 550$ K. The maximum power delivered in this mode of operation is 81 W·m$^{-2}$.

To stablish a reference for later comparison with the experimental results, we note that the limiting radiative reverse saturation current for InAs, $J_{01}$, (when back reflector is used) results into 7.6×10$^{-5}$ A·cm$^{-2}$. If emission towards a substrate would be considered instead, this number should be multiplied by $n^2$ resulting in $J_{01} \approx 9.3\times10^{-4}$ A·cm$^{-2}$ when emission towards an InAs substrate ($n \approx 3.5$ [12]) is considered.

Also from the results in Fig. 1 we observe that, to perform properly, InAs TR didoes should not be impacted by breakdown voltages at least up to reverse voltages in the range of –0.3 V.

With these premises, the objectives of this research were: a) to provide an initial manufacturing route for InAs TR diodes by MBE that exhibit a clean diode current-voltage characteristic curve characterized by a flat reverse saturation current and a breakdown-voltage above 0.3 V to serve as initial workbench for future developments; b) to determine how far our initial MBE technology is from InAs operation in the radiative limit.

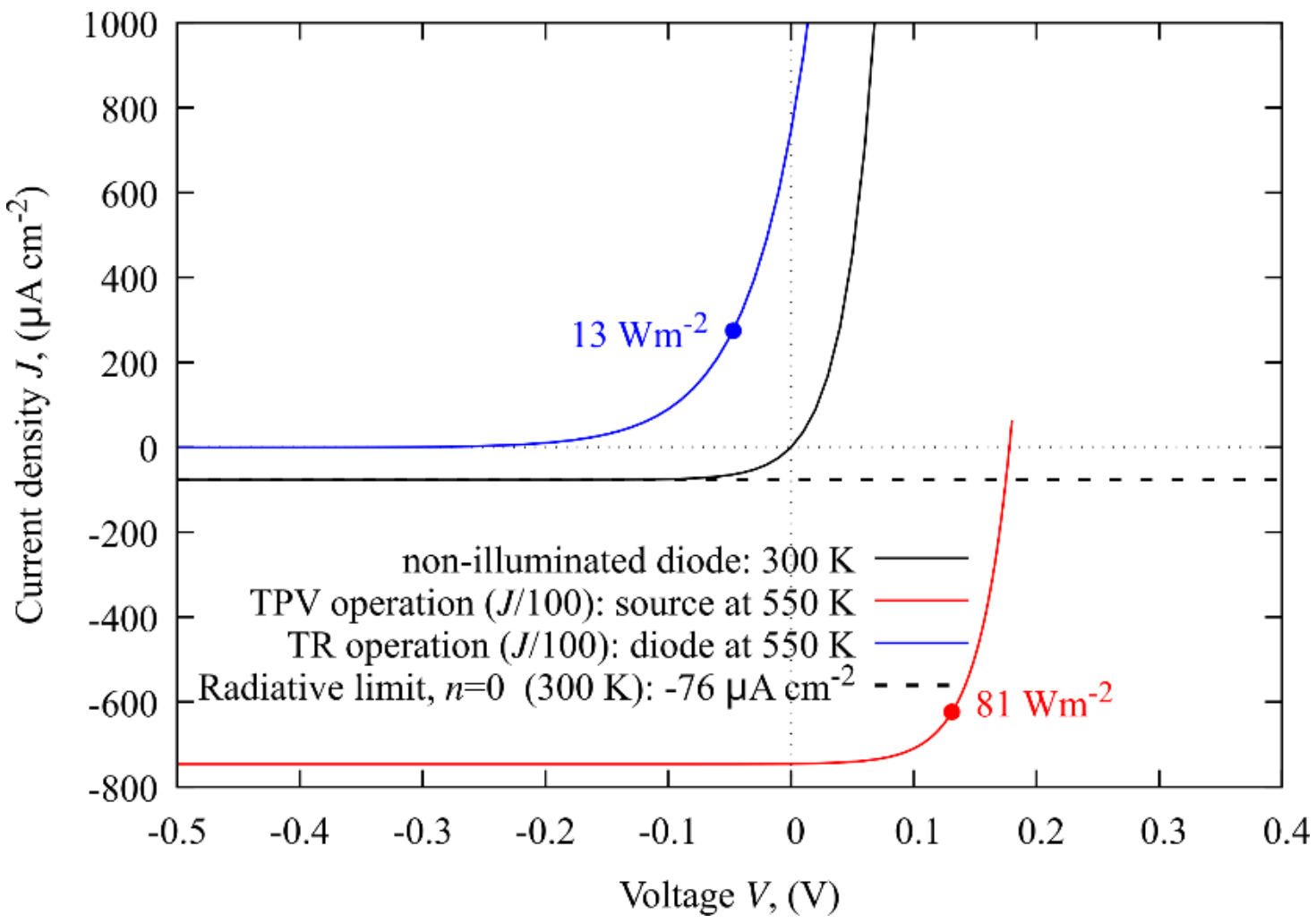

Fig. 1. Simulated current-voltage characteristic of an ideal radiative InAs pn diode operating: non-illuminated (black) with $T_S = T_C = 300$ K; in thermophotovoltaic operation (TPV) when illuminated by a black body at $T_S$ =550 K under maximum concentrated light and the diode at $T_C$ =300 K (red); in thermoradiative (TR) mode with the diode operated at $T_C$ =550 K with the surroundings at $T_S$ =300 K. Notice that the TPV and TR current densities plotted are the result of dividing the actual current density by a factor 100. In this radiative limit, the maximum power obtained in TPV mode is 81 W·m$^{-2}$ and the maximum power in TR mode is 13 W·m$^{-2}$. Voltages refer to the voltage at the p-side of the diode with respect to the n-side. Current is taken as positive when entering through the p-side and exiting from the n-side.

## 3. Description of the diode structures and results

Fig. 2 sketches the four InAs diode structures manufactured by molecular beam epitaxy (MBE) on a Veeco Gen 10. Table I summarizes the MBE growth conditions. All the samples were grown on 2” InAs n$^+$ (1,0,0) substrates, doped with S in the range 1–3×10$^{18}$ cm$^{-3}$ according to their manufacturer (Wafer Technology Ltd.)

For the determination of the temperature of growth, we observed that InAs deoxidation started at 550 ºC according to the thermocouple located close to the wafer molybdenum holder. We will designate the temperature of the thermocouple as $T_{tc}$. Since, according to the literature [13], InAs deoxidation occurs at 530 ºC, we took the opportunity in one of our samples (B15) to calibrate the pyrometer temperature ($T_p$) at this point so that when $T_{tc} = 550$ °C, then we set $T_p$=530 ºC. As indicated in Table I, samples were grown at two different temperatures $T_{tc}$: 490 ºC and 500 ºC. The temperature with the pyrometer could only be monitored for B15. However, since the substrates and holders used were the same for all the samples, we think that the temperature monitored with the pyranometer for B15 can also be assumed to be the same for the other samples. At all the growth temperatures, 2×4 surface reconstruction was observed. For complete deoxidation, substrate temperature was increased 20 ºC remaining at this temperature for about 10 minutes while the $As_2$ valve was fully opened.

For the determination of the growth rate, standard RHHED oscillations [14] were measured. Consistent results were obtained along the (001), (0,–1,1) and (0,1,1) directions. To maximize the number of oscillations, first, during InAs growth with $As_2$ overpressure, the $As_2$ flux was reduced until the onset of 4×2 surface reconstruction was observed. The $As_2$ flux at this onset, $\phi_{As_2,st}$, can be taken as a system independent reference since it is supposed to mark the stoichiometry point at which the flux of As atoms equals the flux of In atoms plus the flux of As atoms that are evaporated from the substrate. Then the In flux was stopped and the surface was allowed to flatten at the temperature of growth for about 5 minutes with an $As_2$ flux around 1.3 times larger than the one measured for stoichiometry. The increment in $As_2$ flux was monitored with an ion-gauge beam flux monitor. The $As_2$ flux was then increased to around 3 times stoichiometry and the In shutter was opened. Oscillations were then registered. For setting this procedure we acknowledge the practical recommendations by F.Bastiman in his blog [15].

After growth, 1×1 $mm^2$ diodes were manufactured. Diode area was defined by mesa etching in $3H_3PO_4$:$4H_2O_2$:$1H_2O$ at room temperature controlling the depth of the etching with a profilometer until the substrate was reached. Etching time was in the order of 1 min for all samples. All the samples were metallized at the top by Joule evaporated gold achieving a total thickness of 2800 Å. Gold contacts were not annealed since we expected reasonable contact quality, and we did not wish to introduce the possibility of gold diffusing deeper in the diode structure and the subsequent incertitude in the interpretation of the measurements at this stage of the research. The substrate was not metalized since, due to the n nature of the substrate and low gap of InAs, reasonable contact quality for our measurements was expected, as indeed it ended up being the case. On the other hand, since the goal of our research is to explore the diode reverse saturation to compare with its radiative limit, contact optimization was not our priority.

The best current density – voltage characteristics of each batch are plotted in Fig. 3.

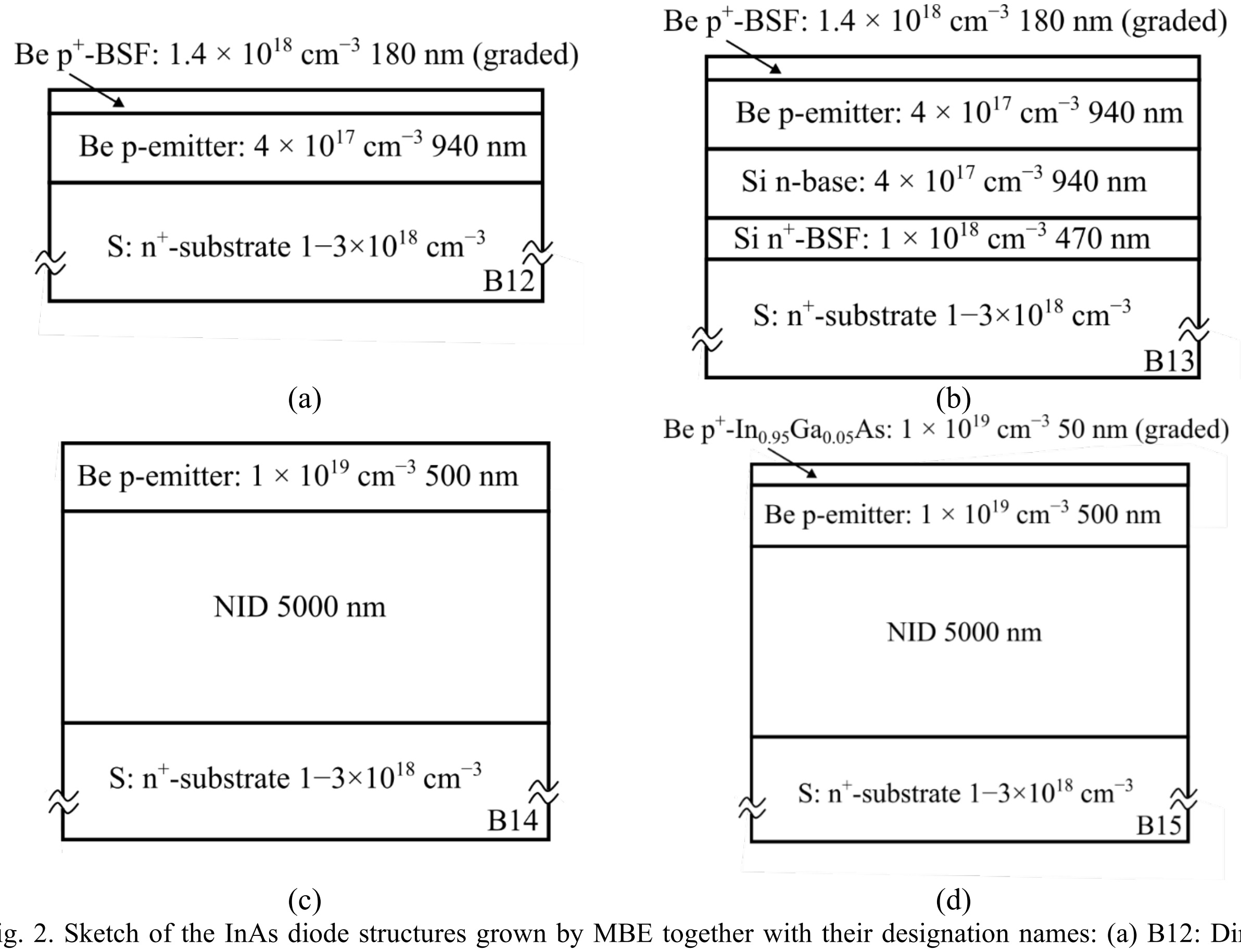

Fig. 2. Sketch of the InAs diode structures grown by MBE together with their designation names: (a) B12: Direct growth of the p emitter on the substrate and BSF layer at the top for surface passivation; (b) B13: growth of p emitter on top of n silicon doped n base. BSF layer at the top and bottom; c) B14: pin structure with highly doped emitter (structure motivated by the study by [16]; d) Same than B14 but with top $In_{0.95}Ga_{0.05}As$ graded layer for passivation purposes and inversion layer mitigation.

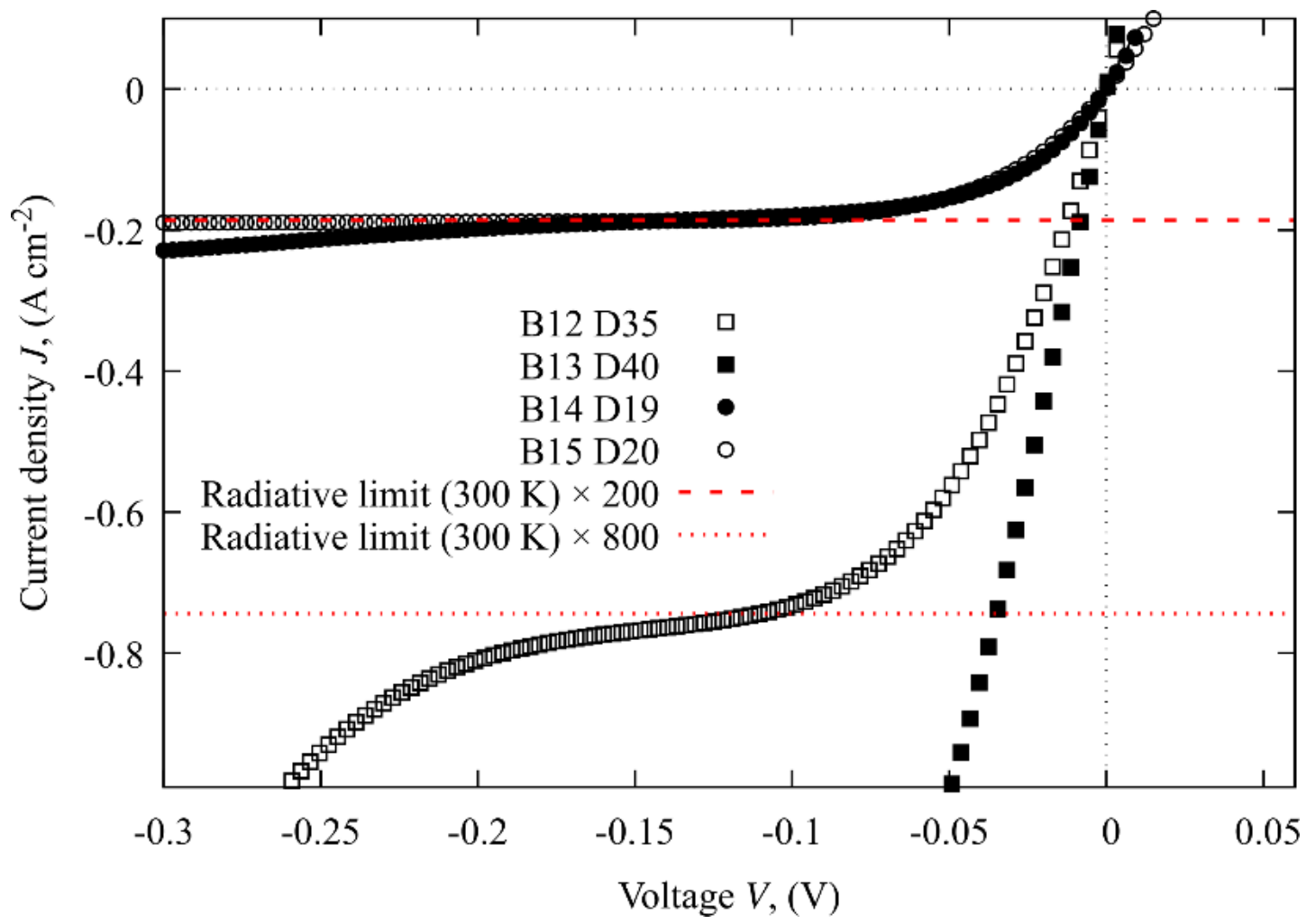

Fig. 3. Best diode current density-voltage characteristic (defined by such that shows lower reverse saturation current density) for each batch. The reverse current-density limit for an InAs diode with back reflector is also shown for reference (dashed line). Measurements at 300 K. Diode area: 1×1 mm$^2$. Comparison with the radiative limit refers to the case of emission towards a substrate for which case the limit is $J_{01} \approx 9.3\times10^{-4}$ A·cm$^{-2}$.

Table I. Growth conditions of the samples studied in this work, including the ratio of their reverse saturation current density to the reverse radiative current density of InAs (9.3×10$^{-4}$ Acm$^{-2}$ for a diode with emission to the substrate).

| Sample | $T_{tc}/T_p$ growth (ºC) | $As_2$ flux $/\phi_{As_2,st}$ | InAs Growth rate (nm·hr$^{-1}$) | In flux (at cm$^{-2}$s$^{-1}$) | Reverse saturation current (mA·cm$^{-2}$) | Reverse saturation current to radiative limit ratio (300 K) |
|---|---|---|---|---|---|---|
| B12 | 500/460 | 3.3 | 470 | 2.3×10$^{14}$ | 750 | 800 |
| B13 | 490/450 | 2.5 | 470 | 2.3×10$^{14}$ | - | - |
| B14 | 490/450 | 2.9 | 330 | 1.6×10$^{14}$ | 190 | 200 |
| B15 | 490/450 | 3.1 | 330 | 1.6×10$^{14}$ | 190 | 200 |

Sample B12 corresponds to the direct growth of a p-InAs layer on top of the n-type substrate. The structure includes a graded top p$^+$ back surface field (BSF) layer aiming to reduce surface recombination and facilitate metal contact. The simplicity of this structure aimed to serve an initial reference. Since, for example, no buffer layer was included in the structure, a current density far from the radiative limit was expected due to the impact of the non-radiative recombination from the substrate. Diodes showed a reverse saturation current 800 times larger than the calculated radiative limit considering the case of a diode that emits to a substrate at the back surface. The diodes also showed a too low breakdown voltage ($\sim -0.2$ V).

Sample B13 aimed to use our own grown n-InAs layer instead of the substrate. Silicon was used as dopant to a nominal doping of 4×10$^{17}$ cm$^{-3}$. However, in this case, no current rectification was observed for the samples and their characteristic appeared nearly short-circuited. We do not have a clear cause for this. We suspected that could be due to a too low $As_2$ flux that could have increased the density of defects and/or the natural presence of an n-type surface inversion layer that has been reported [16] difficult to compensate by means of p doping.

Sample B14 (pin structure) aimed to replicate one of the successful diode structures reported by Ikossi [16]. On the one hand, the non-intentional doped (NID) layer of the structure aims to increase the breakdown voltage. On the other hand, the increased level of doping of the p emitter aims to further facilitate the metal contact and to accentuate the compensation in case of the presence of an inversion layer. Diodes showed improved characteristic with a reverse saturation current 200× the one expected from the radiative limit. In order to finish the surface of the structure in a repetitive and controlled way, immediately after the growth was finished, a 4×4 surface reconstruction was created by decreasing the substrate temperature while preserving $As_2$ flow. The reconstruction was monitored by RHEED and, once the reconstruction was observed, the temperature was decreased further, the $As_2$ flux closed at 300 C and when the sample reached 40 C, the $As_2$ flow was restarted until the RHEED revealed the formation of an amorphous layer of As on the surface. This surface termination was also carried out in the next sample, B15.

In sample B15, also with pin structure, we aimed to reduce surface recombination and to reduce the possible impact of negative charge accumulation at the surface. To this end, we grow a 50-nm graded $p^+$–$In_{0.95}Ga_{0.05}As$ at the top of the structure. The reduced thickness of this layer aimed to accommodate the strain between InAs and $In_{0.95}Ga_{0.05}As$ due to differences in lattice constants between both materials. The samples marginally improved the current density-voltage characteristic of the nearly identical B14 structure. We observed, however, a lower leakage current density. Although this could be due to the presence of the $In_{0.95}Ga_{0.05}As$, we suspect that it results from the fact that in this growth we increased (from 100 ºC to 150 ºC) the temperature difference between the tip and base of the In dual filament Veeco effusion cells, what might have contributed to the reduction of the oval defect density. The reverse saturation current of this batch was also 200× the radiative limit and 10× the value reported by Selvidge et al. for InAs thermophotovoltaic cells (20 mA $cm^{-2}$) [7].

## 4. Conclusions

The InAs diodes grown by MBE with pin structure provide a reasonable starting point for developing TR diodes for applications in which electricity is obtained from heat sources. These structures, particularly B15, show a flat reverse saturation current–voltage characteristic, with negligible leakage current and breakdown voltage higher than 0.3 V. The factors that might have contributed to this result, beside growth parameters such as temperature and $As_2$/In flux ratio are: the termination in a 4×4 surface reconstruction an thin As amorphous layer after growth, the finalization of the growth with a thin graded $p^+$–$In_{0.95}Ga_{0.05}As$ layer and the increase in the temperature difference between tip and base of the In effusion cell from 100 ºC to 150 ºC. The reverse saturation current measured is still 200× the radiative limit of an ideal thermoradiative diode emitting towards the substrate. Further improvement should be investigated by sweeping the growth parameters and implementing a diode structure with radiative emission to air instead to substrate.

## 5. References

[1] I. Vurgaftman, J. R. Meyer, and L. R. Ram-Mohan, "Band parameters for III–V compound semiconductors and their alloys," *J. Appl. Phys.*, vol. 89, no. 11, pp. 5815–5875, Jun. 2001, doi: 10.1063/1.1368156.

[2] D. L. Rode, "Electron Mobility in Direct-Gap Polar Semiconductors," *Phys. Rev. B*, vol. 2, no. 4, pp. 1012–1024, 1970, doi: 10.1103/PhysRevB.2.1012.

[3] D.-H. Kim and J. A. del Alamo, "30-nm InAs pseudomorphic HEMTs on an InP substrate with a current-gain cutoff frequency of 628 GHz," *IEEE Electron Device Lett.*, vol. 29, no. 8, pp. 830–833, 2008, doi: 10.1109/LED.2008.2000794.

[4] J. Heremans, "Solid state magnetic field sensors and applications," *J. Phys. Appl. Phys.*, vol. 26, no. 8, pp. 1149–1168, 1993, doi: 10.1088/0022-3727/26/8/001.

[5] S. Maimon and G. W. Wicks, "nBn detector, an infrared detector with reduced dark current and higher operating temperature," *Appl. Phys. Lett.*, vol. 89, no. 15, 2006, doi: 10.1063/1.2360235.

[6] Q. Lu *et al.*, "InAs thermophotovoltaic cells with high quantum efficiency for waste heat recovery applications below 1000 °C," *Sol. Energy Mater. Sol. Cells*, vol. 179, pp. 334–338, Jun. 2018, doi: 10.1016/j.solmat.2017.12.031.

[7] J. Selvidge, R. M. France, J. Goldsmith, P. Solanki, M. A. Steiner, and E. J. Tervo, “Large Area Near-Field Thermophotovoltaics for Low Temperature Applications,” *Adv. Mater.*, vol. 37, no. 5, p. 2411524, 2025, doi: 10.1002/adma.202411524.

[8] R. Strandberg, “Theoretical efficiency limits for thermoradiative energy conversion,” *J. Appl. Phys.*, vol. 117, no. 5, p. 055105, Feb. 2015, doi: 10.1063/1.4907392.

[9] W. Shockley and H. J. Queisser, “Detailed Balance Limit of Efficiency of p-n Junction Solar Cells,” *J. Appl. Phys.*, vol. 32, no. 3, pp. 510–519, 1961.

[10] G. L. Araujo and A. Marti, “Absolute limiting efficiencies for photovoltaic energy conversion,” *Sol. Energy Mater. Sol. Cells*, vol. 33, no. 2, pp. 213–240, 1994.

[11] A. Martí, J. L. Balenzategui, and R. F. Reyna, “Photon recycling and Shockley’s diode equation,” *J. Appl. Phys.*, vol. 82, no. 8, pp. 4067–4075, 1997.

[12] S. Adachi, “Optical dispersion relations for GaP, GaAs, GaSb, InP, InAs, InSb, $Al_{x}Ga_{1-x}As$, and $In_{1-x}Ga_{x}As_{y}P_{1-y}$,” *J. Appl. Phys.*, vol. 66, no. 12, pp. 6030–6040, Dec. 1989, doi: 10.1063/1.343580.

[13] H. Ye *et al.*, “MBE growth optimization of InAs (001) homoepitaxy,” *J. Vac. Sci. Technol. B Nanotechnol. Microelectron. Mater. Process. Meas. Phenom.*, vol. 31, no. 3, p. 03C135, May 2013, doi: 10.1116/1.4804397.

[14] W. Braun, *Applied RHEED: reflection high-energy electron diffraction during crystal growth*. in Springer tracts in modern physics, no. v. 154. Berlin ; New York: Springer, 1999.

[15] F. Bastiman, “Little known MBE facts,” MBE Blog (accessed 2025). [Online]. Available: https://faebianbastiman.wordpress.com/mbe-blog-contents-page/

[16] K. Ikossi, “InAs Device Process Development and Characterization:,” Defense Technical Information Center, Fort Belvoir, VA, May 2003. doi: 10.21236/ADA413747.

.